# Title: Discovery of Phylogenetic Relevant Y-chromosome Variants in 1000 Genomes Project Data

Authors: Chuan-Chao Wang[1], Hui Li[1, *]

**Affiliations**
1. State Key Laboratory of Genetic Engineering and MOE Key Laboratory of Contemporary Anthropology, School of Life Sciences and Institutes of Biomedical Sciences, Fudan University, Shanghai 200433, China
* Correspondence to: lihui.fudan@gmail.com

**Abstract**
Current Y chromosome research is limited in the poor resolution of Y chromosome phylogenetic tree. Entirely sequenced Y chromosomes in numerous human individuals have only recently become available by the advent of next-generation sequencing technology. The 1000 Genomes Project has sequenced Y chromosomes from more than 1000 males. Here, we analyzed 1000 Genomes Project Y chromosome data of 1269 individuals and discovered about 25,000 phylogenetic relevant SNPs. Those new markers are useful in the phylogeny of Y chromosome and will lead to an increased phylogenetic resolution for many Y chromosome studies.

## Introduction

The paternally inherited Y chromosome has been widely used in anthropology and population genetics to understand origin and migration of human populations[1]. With a very low mutation rate on the order of $3.0 \times 10^{-8}$ mutations/nucleotide/generation[2], single nucleotide polymorphisms (SNPs) of Y chromosome have been used in constructing a phylogenetic tree linking all the Y chromosome lineages from world populations[3, 4]. Since the middle of 1980s, Y-specific probes had been isolated from cosmid libraries and used in association with a set of restriction enzymes to search for male specific restriction fragment length polymorphisms (RFLPs)[5-9]. Since the late 1990s, denaturing high-performance liquid chromatography (DHPLC) method has been used to detect the SNPs in the single-copy regions of MSY[10, 11]. During the last ten years, a robust genealogical tree of human Y chromosomes based on about three thousand stable SNPs has been built, permitting inference of human population demographic history[12, 13]. However, current Y chromosome research is still limited in the poor resolution for some specific Y chromosome branches, such as haplogroup C-M130, D-M174, N-M231, O-M175, H-M69, and L-M11. Despite the huge population of those haplogroups, there have been fewer markers defined in those haplogroups than in haplogroups R and E. For instance, three Y-SNP markers, 002611, M134 and M117, represent about 260 million people in East Asia, but downstream markers are far from enough to reveal informative genetic substructures of those populations[1]. Entirely sequenced Y chromosomes in numerous human individuals have only recently become available by the advent of next-generation sequencing technology[14, 15, 16, 17]. For instance, the 1000 Genomes Project has sequenced Y chromosomes from more than 1000 males. Here, we analyzed 1000 Genomes Project Y chromosome data of 1269 individuals and discovered thousands of new SNPs that might be useful in the phylogeny of Y chromosome. Those new markers will lead to an increased phylogenetic resolution for many Y chromosome studies.

## Materials and Methods

The phylogenetic tree was based on ISOGG at 6 September 2013 (http://www.isogg.org/). SAMtools (version 0.1.9) view was used to download mapped bam files from publicly accessible FTP sites at the European Bioinformatics Institute (ftp://ftp.1000genomes.ebi.ac.uk/vol1/ftp/) and the National Center for Biotechnology Information (ftp://ftp-trace.ncbi.nih.gov/1000genomes/ftp/). Reads that were uniquely mapped on Y chromosome with a quality $\geqslant 15$ were extracted from sam files and transformed into bam files with SAMtools[18]. Duplicates were removed by samtools rmdup. Variations were called by SAMtools mpileup[18]. The resulting BCF file was then converted into VCF format by using the bcftools[18]. Haplogroups were classified by using the WHY.pl and AMY-tree.pl scripts[19]. To evaluate the accuracy of haplogroup assignment, maximum likelihood haplogroup trees using the HKY85 model were produced by PhyML (version 20120412)[20], and bootstrap values were produced using 100 subsamplings. Heterozygous calls and calls with phred-scaled quality <30 were removed in constructing the trees. Taken the tree topology into consideration, the VCF files were opened in MS Excel for visual identification of potential phylogenetic relevant SNPs. Novel variants were filtered by verifying that all other haplogroup control samples bore the ancestral allele, and by identifying at least two samples in the case haplogroup that carried the same derived allele.

## Results

Overall about 25,000 SNPs were identified that has not been publicly named before (http://www.isogg.org/). We designated each of these SNP a name beginning with 'F' (for Fudan University) and the initials of each haplogroup.

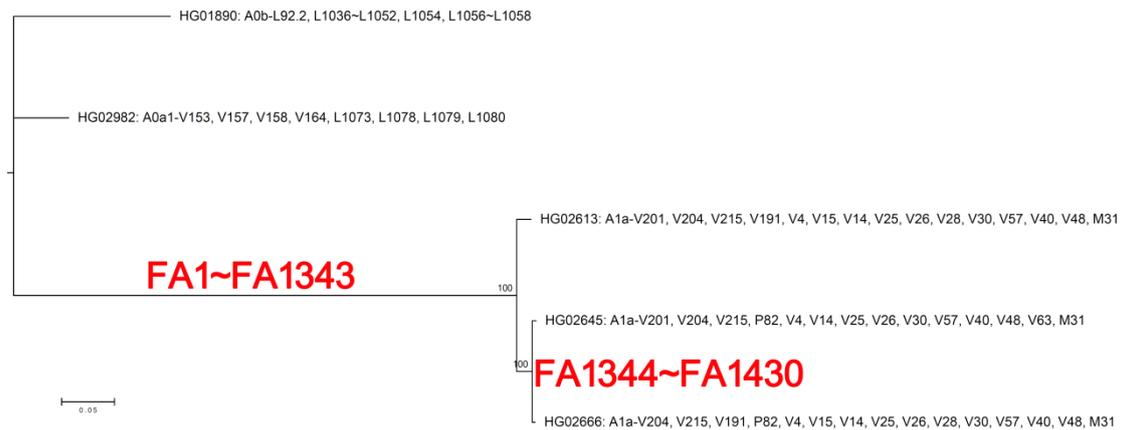

Figure 1. The maximum likelihood tree of haplogroup A.

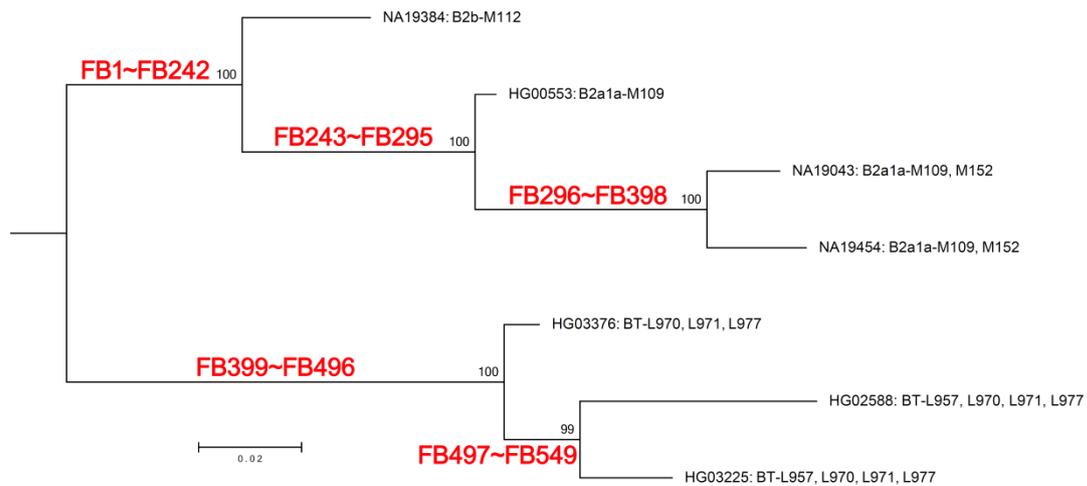

Figure 2. The maximum likelihood tree of haplogroup B.

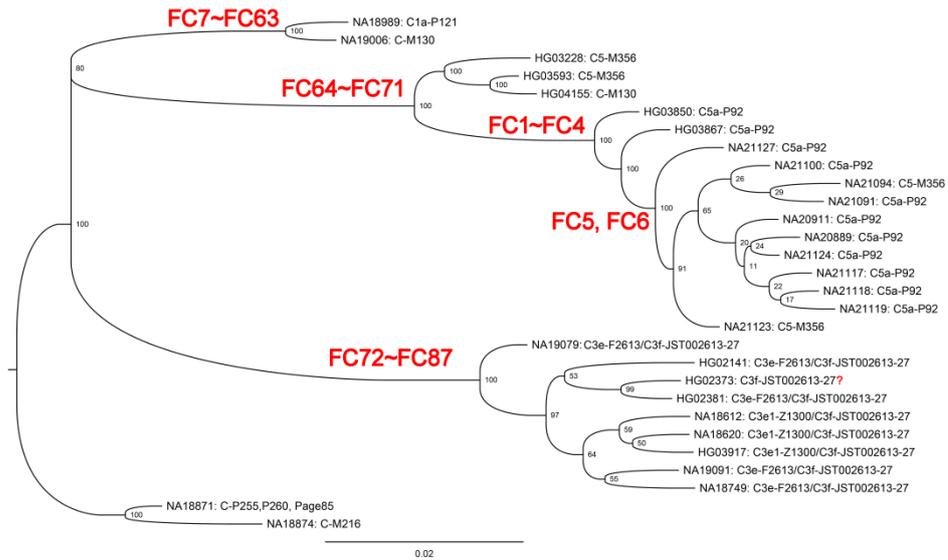

Figure 3. The maximum likelihood tree of haplogroup C.

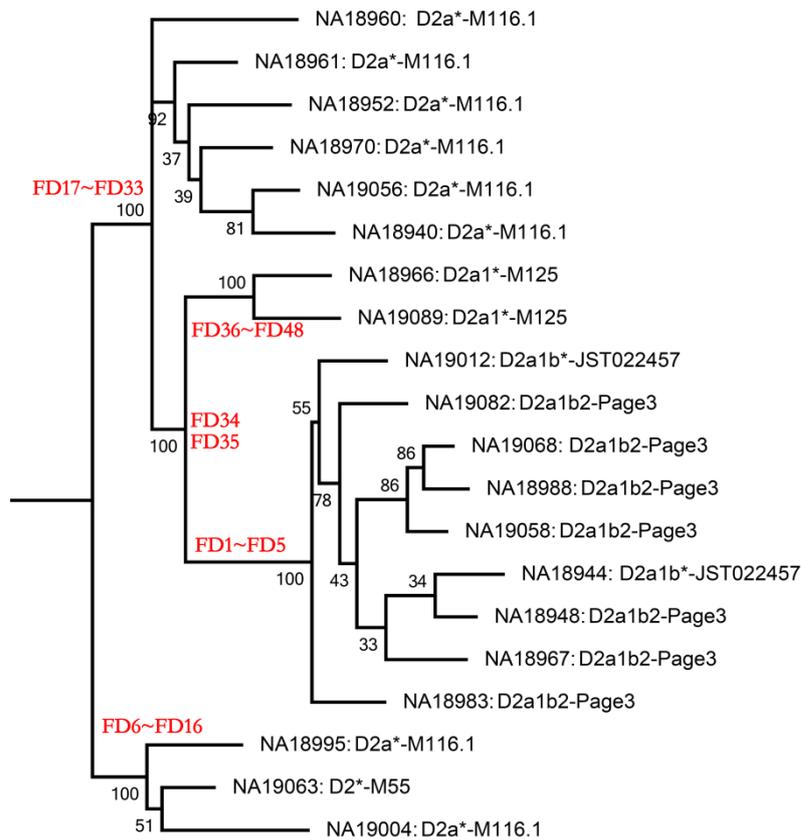

Figure 4. The maximum likelihood tree of haplogroup D.

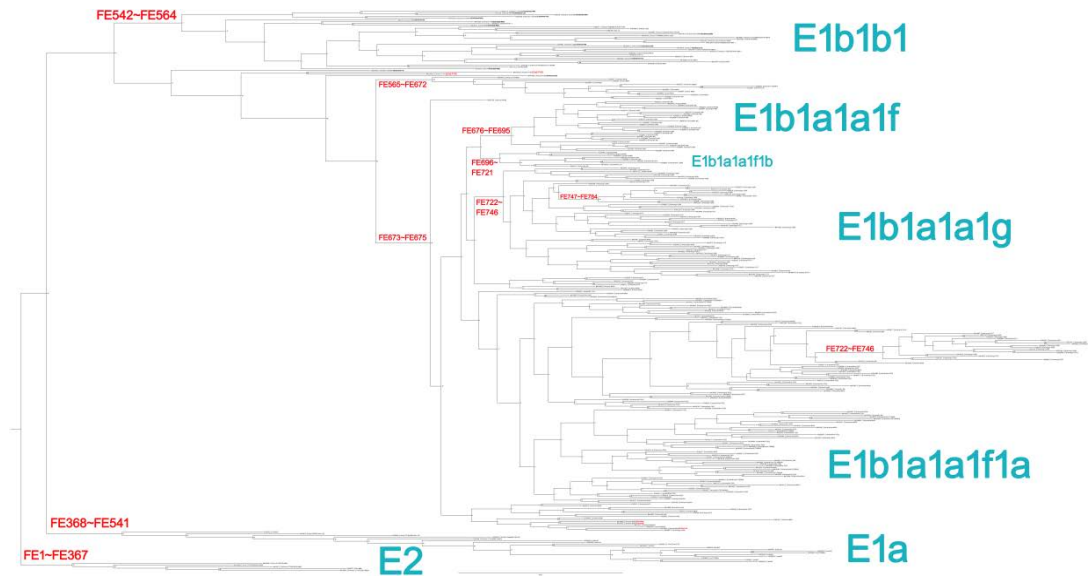

Figure 5. The maximum likelihood tree of haplogroup E.

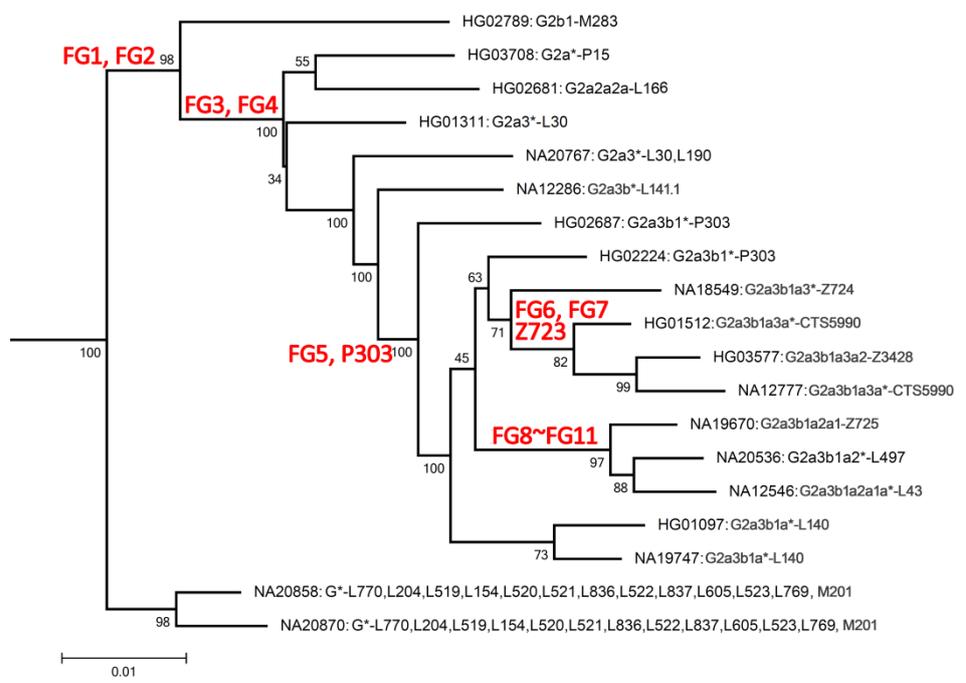

Figure 6. The maximum likelihood tree of haplogroup G.

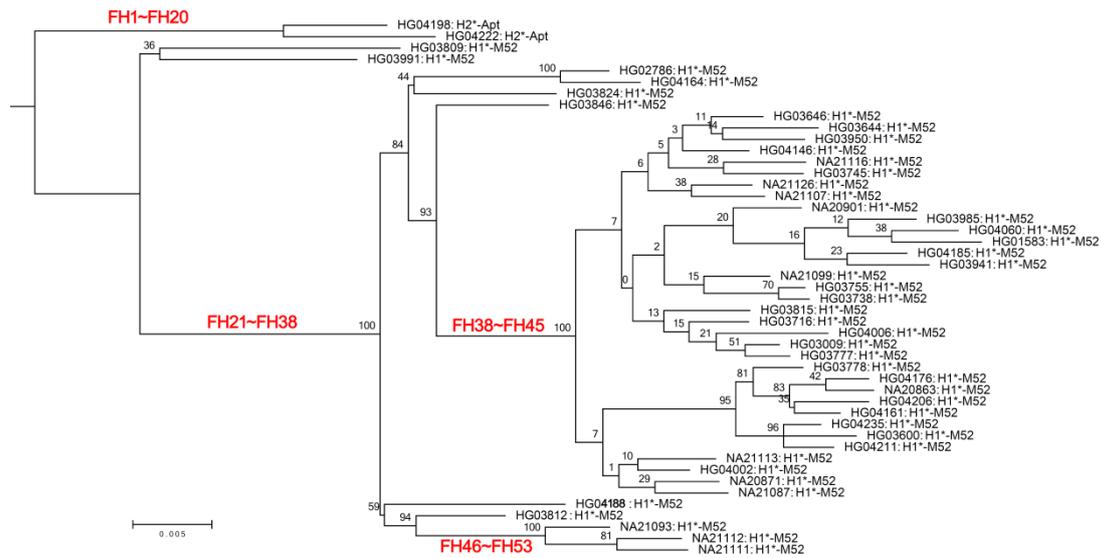

Figure 7. The maximum likelihood tree of haplogroup H.

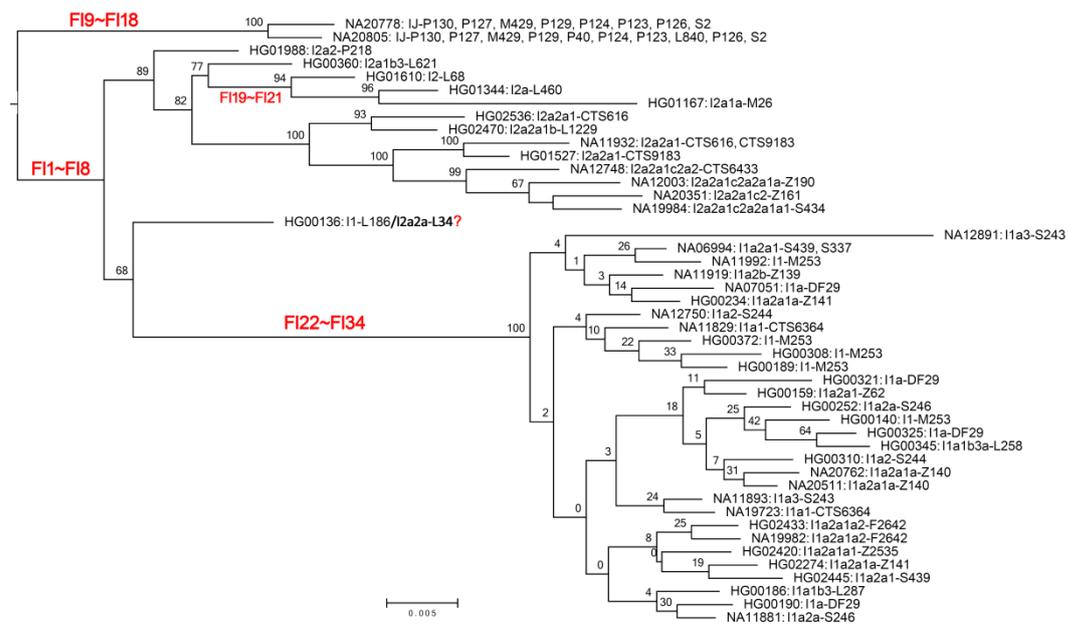

Figure 8. The maximum likelihood tree of haplogroup I.

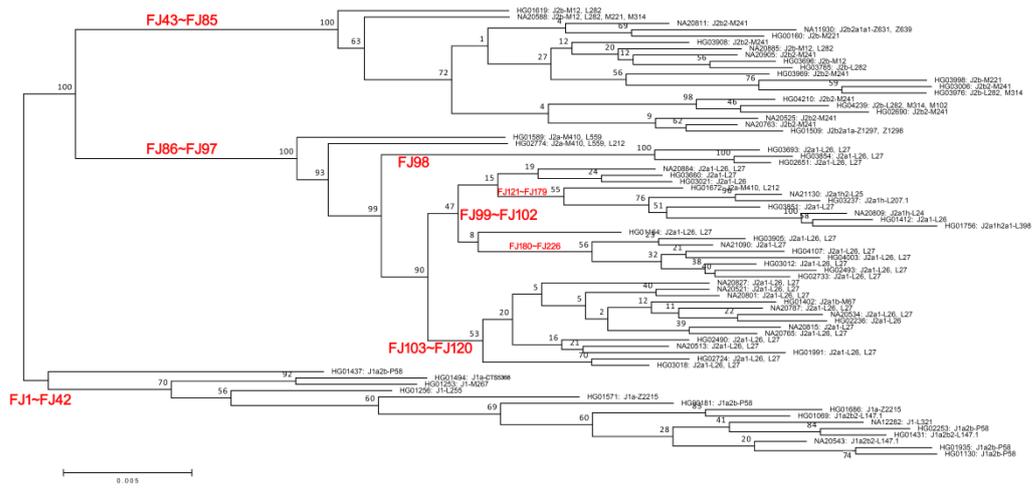

Figure 9. The maximum likelihood tree of haplogroup J.

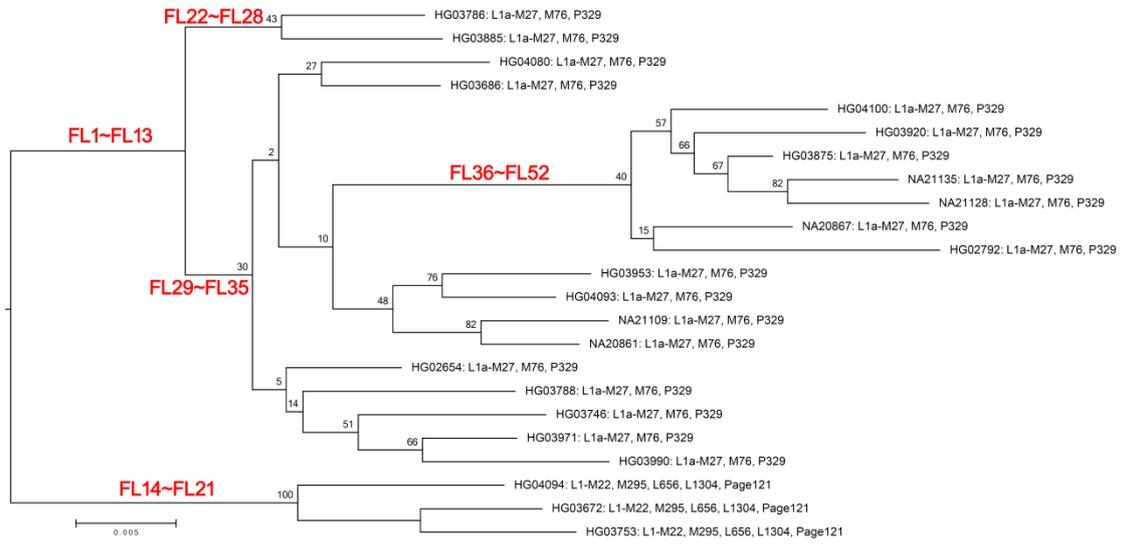

Figure 10. The maximum likelihood tree of haplogroup L.

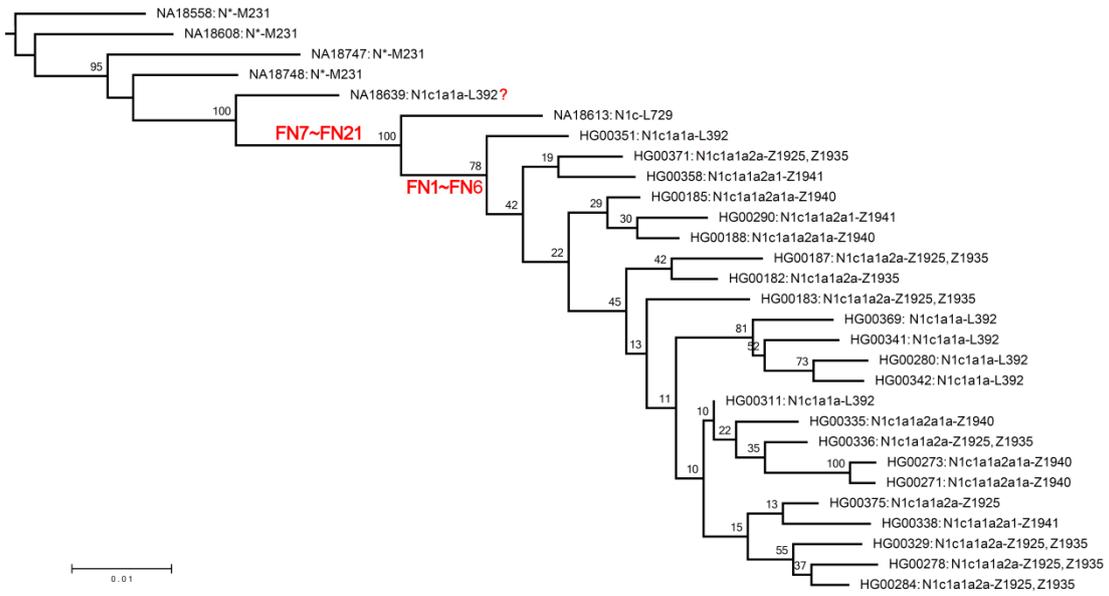

Figure 11. The maximum likelihood tree of haplogroup N.

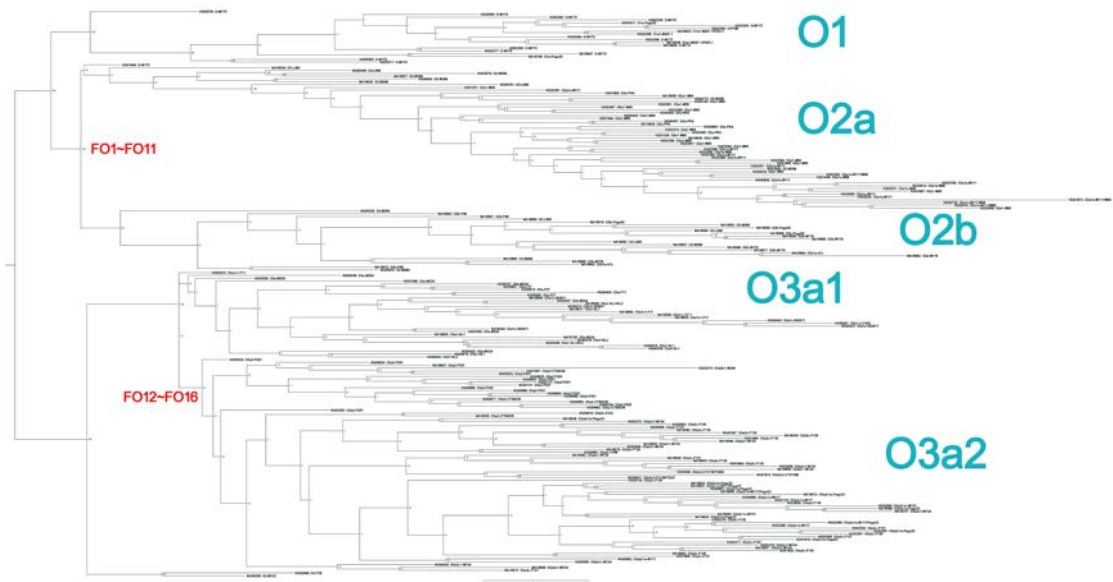

Figure 12. The maximum likelihood tree of haplogroup O.

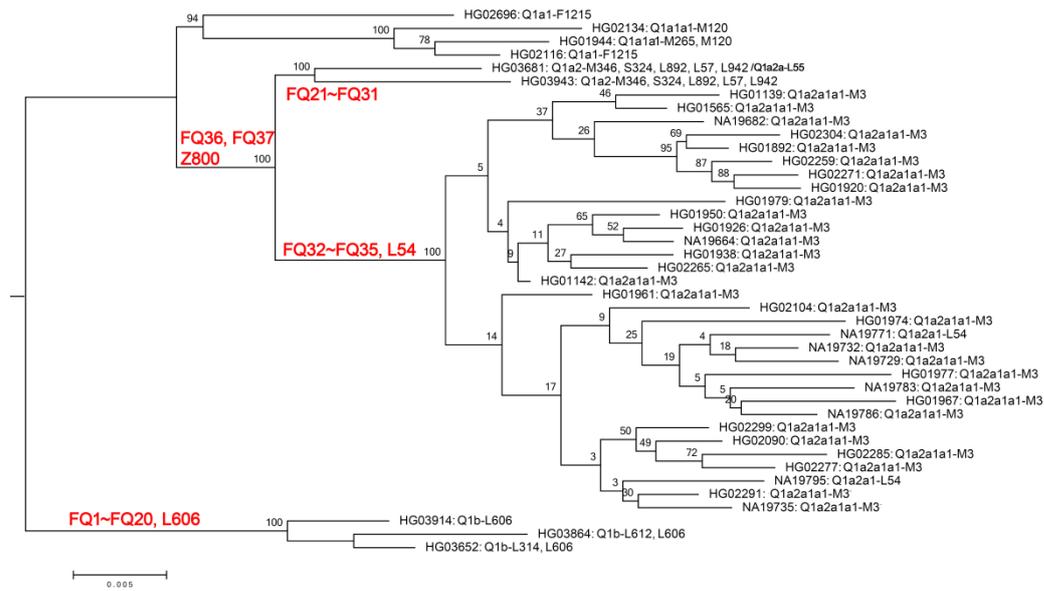

Figure 13. The maximum likelihood tree of haplogroup Q.

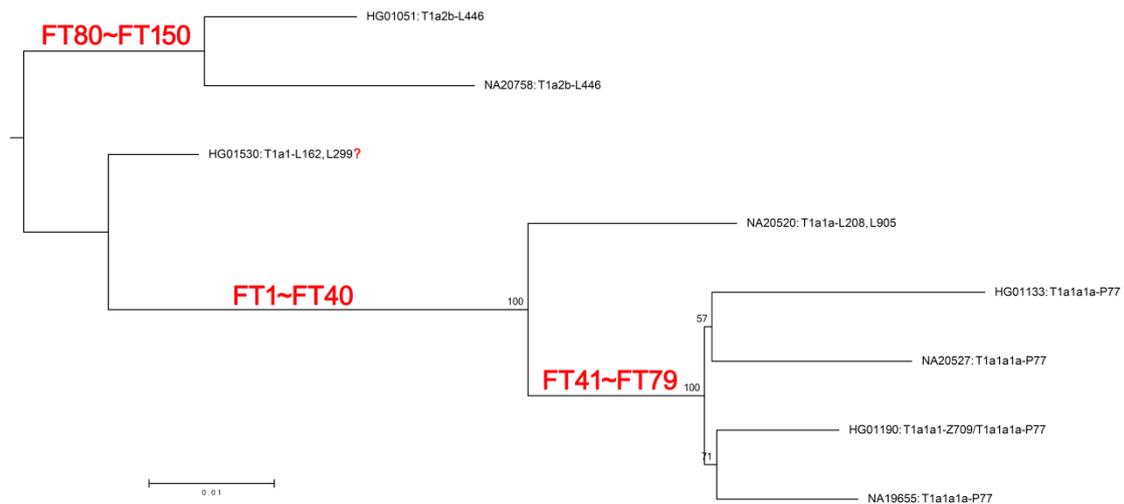

Figure 14. The maximum likelihood tree of haplogroup T.

**Acknowledgments**

This work was supported by the National Excellent Youth Science Foundation of China (31222030), National Natural Science Foundation of China (31071098, 91131002), Shanghai Rising-Star Program (12QA1400300), Shanghai Commission of Education Research Innovation Key Project (11zz04), and Shanghai Professional Development Funding (2010001).

**Supplementary Material**

Genotypes of each sample, newly discovered phylogenetic relevant SNPs, and high resolution figures: http://comonca.org.cn/Y_tree/